\documentclass[final]{svjour2}
\usepackage{graphicx}
\usepackage{rotating}
\usepackage{amssymb}
\usepackage{mathptmx}
\usepackage[numbers]{natbib}
\usepackage{amsmath,amsfonts,amssymb}
\makeatletter
\journalname{Journal of Low Temperature Physics}

\bibpunct{}{}{,}{s}{}{,}

\begin{document}

\newcommand{\hdblarrow}{H\makebox[0.9ex][l]{$\downdownarrows$}-}
\title{On reflectionless nature of self-consistent multi-soliton solutions in Bogoliubov--de~Gennes and chiral Gross--Neveu~models}

\author{Daisuke A. Takahashi$^{1,2}$ \and Muneto Nitta$^{1,3}$}

\institute{
1:Research and Education Center for Natural Sciences, Keio University, Hiyoshi 4-1-1, Yokohama, Kanagawa 223-8521, Japan \\
2: Department of Basic Science, The University of Tokyo, Tokyo, 153-8902, Japan \\ 
3: Department of Physics, Keio University, Hiyoshi 4-1-1, Yokohama, Kanagawa 223-8521, Japan \\
\email{takahashi@vortex.c.u-tokyo.ac.jp}
}

\date{\today}

\maketitle


\begin{abstract}
Recently the most general static self-consistent multi-soliton solutions in Bogoliubov-de Gennes and chiral Gross-Neveu systems are derived by the present authors [D. A. Takahashi and M. Nitta, Phys. Rev. Lett. \textbf{110}, 131601 (2013)]. Here we show a few complementary results, which were absent in the previous our work. We prove \textit{directly from the gap equation} that the self-consistent solutions need to have reflectionless potentials. We also give the self-consistent condition for the system consisting of only right-movers, which is more used in high-energy physics.

\end{abstract}

\section{Introduction}
\indent The Bogoliubov--de Gennes (BdG) equation and the gap equation describe spatially inhomogeneous states in various kinds of condensed matter systems, such as superconductors\cite{DeGennes}, polyacetylene\cite{TakayamaLinLiuMaki}, and ultracold atomic Fermi gases. The equivalent equations also appear in the mean field theory of the Nambu--Jona-Lasinio (NJL) or the chiral Gross--Neveu (GN) model in high-energy physics\cite{NambuJonaLasinio,GrossNeveu,DashenHasslacherNeveu}.\\
\indent It is generally a difficult problem to obtain a self-consistent exact solution satisfying not only 
the BdG equation but also the gap equation, and only a few analytic examples were known so far such as the one- and two-kink (polaron in polyacetylene) solutions \cite{TakayamaLinLiuMaki,DashenHasslacherNeveu,Shei,OkunoOnodera,Feinberg} and the real\cite{BrazovskiiGordyuninKirova,Horovitz} and complex\cite{BasarDunne} kink-crystals. 

Recently, the present authors have determined the most general self-consistent solutions under uniform boundary conditions\cite{TakahashiNitta}. The solutions describe $ n $-soliton states, in which their positions are arbitrary but their phase shifts must be discretized. More recently, these solutions have been generalized to the time-dependent case\cite{DunneThies}.

\indent In this paper, we show several complementary results, which were absent in the previous our work. First, we prove \textit{directly from the gap equation} that the self-consistent soliton solutions need to have reflectionless potentials, using the form of Jost solutions derived from the Riemann-Hilbert approach\cite{AblowitzSegur}. In the preceding works, this fact was proved by the functional derivative with respect to the reflection coefficient\cite{DashenHasslacherNeveu,Shei}. Our derivation will be more comprehensive because we can see directly how the reflection coefficient vanishes.  Next, we give the self-consistent condition of the system consisting of only right-movers, which is more used in high-energy physics. The resultant condition is consistent with the time-independent case of the recent work\cite{DunneThies}.

\section{Model}
	\begin{figure}[t]
		\begin{center}
		\includegraphics{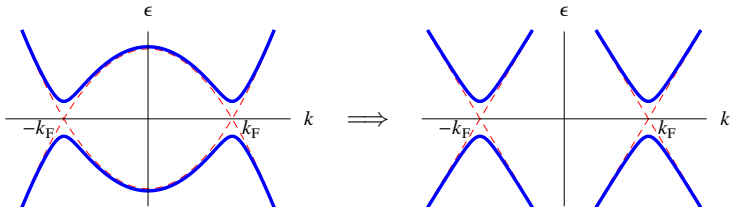}
		\caption{\label{fig:andapp}(Color online) Schematic of the Andreev approximation. The left (right) figure shows the dispersion relation before (after) the Andreev approximation. The red dashed line represents the dispersion relation of free particles and holes and the blue solid line represents the dispersion relation of quasiparticles in the gapped system.}
		\end{center}
	\end{figure}
	The one-dimensional BdG system is given by the BdG equation and the gap equation as a self-consistent condition
	\begin{align}
		\begin{pmatrix} -\frac{1}{2}\partial_x^2-\mu_\uparrow & \Delta(x) \\ \Delta(x)^* & \frac{1}{2}\partial_x^2+\mu_\downarrow \end{pmatrix}\begin{pmatrix}u \\ v \end{pmatrix}=\epsilon \begin{pmatrix} u \\ v \end{pmatrix},\quad -\frac{\Delta(x)}{g}=\sum_{\text{occupied states}} uv^*,
	\end{align}
	where $ \mu_{\uparrow, \downarrow}=\frac{k_F^2}{2}\pm h $ with a Fermi momentum $ k_F $ and a magnetic field $ h $. In this paper we only consider the case $ h=0 $. Following the Andreev approximation, we linearize the dispersion relation around the Fermi points by substituting $ (u(x),v(x))=\mathrm{e}^{\pm\mathrm{i}k_Fx}(u_{\text{R,L}}(x),v_\text{R,L}(x)) $ near the right and left Fermi point $ k=\pm k_F $ and ignoring second-order derivatives (See Fig.~\ref{fig:andapp}). Moving to the new unit with $ k_F=1 $, we obtain the BdG equation for the right- and left-movers
	\begin{align}
		\begin{pmatrix} -\mathrm{i}\partial_x & \Delta(x) \\ \Delta(x)^* & \mathrm{i}\partial_x \end{pmatrix}\begin{pmatrix} u_{\text{R}} \\ v_{\text{R}} \end{pmatrix}=\epsilon \begin{pmatrix} u_{\text{R}} \\ v_{\text{R}} \end{pmatrix},\quad \begin{pmatrix} \mathrm{i}\partial_x & \Delta(x) \\ \Delta(x)^* & -\mathrm{i}\partial_x \end{pmatrix}\begin{pmatrix} u_{\text{L}} \\ v_{\text{L}} \end{pmatrix}=\epsilon \begin{pmatrix} u_{\text{L}} \\ v_{\text{L}} \end{pmatrix} \label{eq:BdGLR01}
	\end{align}
	and the gap equation
	\begin{align}
		-\frac{\Delta(x)}{g}=\sum_{\text{occupied states}} u_{\text{R}}^{}v_{\text{R}}^*+u_{\text{L}}^{}v_{\text{L}}^*. \label{eq:BdGLR02}
	\end{align}
	If we consider the system consisting of only right-movers, the fundamental equations are given by
	\begin{align}
		\begin{pmatrix} -\mathrm{i}\partial_x & \Delta(x) \\ \Delta(x)^* & \mathrm{i}\partial_x \end{pmatrix}\begin{pmatrix} u \\ v \end{pmatrix}=\epsilon \begin{pmatrix} u \\ v \end{pmatrix},\quad -\frac{\Delta(x)}{g}=\sum_{\text{occupied states}} uv^*, \label{eq:BdGR01}
	\end{align}
	which corresponds to the NJL or the chiral GN model\cite{NambuJonaLasinio,GrossNeveu,DashenHasslacherNeveu}.\\ 
 	\indent Henceforth, when we cite Eq.~(x) from our Letter \cite{TakahashiNitta}, we write it as Eq.~(L.x). Similarly, Eq.~(y) in the Supplemental Material of our Letter \cite{TakahashiNitta} is written as Eq.~(S.y). We assume that the gap function $ \Delta(x) $ satisfies the finite-density boundary condition (S.2). Since the solution of the BdG equation for left-movers is expressed by the one for right-movers [Eq.~(L.21)], we can write down the gap equation only using the quantities of right-movers, and we always do so and omit the subscript R hereafter.\\
 	\indent We use the same definitions of right and left Jost solutions $ f_\pm(x,s) $ and the transition coefficients $ a(s) $ and $ b(s) $ in (S.10)-(S.17), where $ s $ is a uniformizing variable defined by $ \epsilon(s)=\frac{m}{2}(s+s^{-1}) $ and $ k(s)=\frac{m}{2}(s-s^{-1}) $ [Eq.~(L.10)]. In order to fit the notations with Ref.~\cite{TakahashiNitta}, we define the scattering states $ (u(x,s),v(x,s)) $ and bound states $ (u_j(x),v_j(x)) \ (j=1,\dots,n) $  using the left Jost solution as follows:
	\begin{align}
		\begin{pmatrix} u(x,s) \\ v(x,s) \end{pmatrix}:=f_-(x,s^{-1}),\quad \begin{pmatrix} u_j(x) \\ v_j(x) \end{pmatrix}=\begin{pmatrix} f_j(x) \\ s_j f_j(x)^* \end{pmatrix}=-C_jf_-(x,s_j), \label{eq:reful401}
	\end{align}
	where $ s_j (j=1,\dots,n) $ is a discrete eigenvalue satisfying $ |s_j|=1 $, and $ C_j=|b(s_j)|c_j $ is a normalization constant of the $j$-th bound state (See Ref.~\cite{TakahashiNitta} for more detail). We also write $ \kappa_j=-\mathrm{i}k(s_j) $ and $ e_j(x)=C_j\mathrm{e}^{\kappa_j x} $. If $ \Delta(x) $ is a reflectionless potential, they reduce to (L.18) and (L.19) with the linear equation (L.13), or equivalently, (S.83) and (S.84) with (S.79). The asymptotic form of scattering states are given by
	\begin{align}
		\begin{pmatrix} u(x,s) \\ v(x,s) \end{pmatrix} \!\rightarrow\! \begin{cases} \begin{pmatrix}1\\[-0.75ex] s^{-1}\end{pmatrix}\mathrm{e}^{\mathrm{i}k(s)x} &\!\!(x\rightarrow-\infty) \\ \mathrm{e}^{\mathrm{i}\theta\sigma_3}\left[a(s)^* \begin{pmatrix} 1 \\[-0.75ex] s^{-1}\end{pmatrix}\mathrm{e}^{\mathrm{i}k(s)x}-b(s)\begin{pmatrix}s^{-1} \\[-0.75ex] 1 \end{pmatrix}\mathrm{e}^{-\mathrm{i}k(s)x}\right]&\!\!(x\rightarrow+\infty). \end{cases} \label{eq:reful301}
	\end{align}
	The transmission and reflection coefficients are defined by $ t(s)=1/a(s) $ and $ r(s)=b(s)/a(s) $.\\
	\indent For a system with both right- and left-movers described by Eqs.~(\ref{eq:BdGLR01}) and (\ref{eq:BdGLR02}), we consider the same occupation state considered in Ref.~\cite{TakahashiNitta}.   As we derive in the next section, the gap equation in the infinite-length limit is given by
	\begin{align}
		0=&\sum_{\text{b.s.}}\nu_ju_j(x)v_j(x)^*\nonumber \\
		&+\left[\int_{-\infty}^0-\int_0^\infty\right]\frac{\mathrm{d}s}{2\pi}\left( \frac{m}{2}\left( u(x,s)v(x,s)^*+r(s)^*u(x,s)^2 \right)-\frac{\Delta(x)}{2s} \right)\!, \label{eq:gapeqforLR}
	\end{align}
	which gives the generalization of Eq.~(L.27) with a non-vanishing reflection coefficient $ r(s) $. For a system with only right-movers described by Eq.~(\ref{eq:BdGR01}), we also consider the similar occupation state,  in which the negative-energy scattering states are completely filled and positive ones are empty, and the bound states are filled partially. Writing the filling rate of the $ j $-th bound state as $ \nu_j:=N_j/N $, we obtain the following gap equation in the next section:
	\begin{align}
		0=\sum_{\text{b.s.}}\nu_ju_j(x)v_j(x)^*+\int_{-\infty}^0\frac{\mathrm{d}s}{2\pi}\left( \frac{m}{2}\left( u(x,s)v(x,s)^*+r(s)^*u(x,s)^2 \right)-\frac{\Delta(x)}{2s} \right)\!. \label{eq:gapeqforR}
	\end{align}
\section{Gap equation}
	\indent In this section, we derive the gap equation in the infinite-volume limit. In order to avoid the mathematical difficulty of infinite systems, we first treat the finite system with the length $ L $ in the interval $ [-\frac{L}{2},\frac{L}{2}] $, and take the limit $ L\rightarrow\infty $. 
\subsection{Discretized eigenstates in a finite system}
	\indent To include the effect of solitons' phase shifts, we consider the following ``twisted'' periodic boundary condition:
	\begin{align}
		\Delta(\tfrac{L}{2})=\mathrm{e}^{2\mathrm{i}\theta}\Delta(-\tfrac{L}{2}),\quad u(\tfrac{L}{2})=\mathrm{e}^{\mathrm{i}\theta}u(-\tfrac{L}{2}),\quad v(\tfrac{L}{2})=\mathrm{e}^{-\mathrm{i}\theta}v(-\tfrac{L}{2}). \label{eq:reful303}
	\end{align}
	We note, however, that the final expression Eq.~(\ref{eq:gapssinf01}) below does not depend on the detail of the boundary condition. We assume that $ L $ is sufficiently large and hence the asymptotic form of the left Jost solution (\ref{eq:reful301}) can be used in the substitution at $ x=\pm\frac{L}{2} $. 
	Then, after a straightforward calculation, we obtain a set of discretized scattering eigenstates satisfying the boundary condition (\ref{eq:reful303}), which is given by
	\begin{align}
		\begin{pmatrix} u_{\text{PB}}(x,s) \\ v_{\text{PB}}(x,s) \end{pmatrix}\!=\mathrm{e}^{-\mathrm{i}\varphi(s)}\!\begin{pmatrix} u(x,s) \\ v(x,s) \end{pmatrix}\!+\mathrm{e}^{\mathrm{i}\varphi(s)}\!\begin{pmatrix} v(x,s)^* \\ u(x,s)^* \end{pmatrix}\!\!,\ \mathrm{e}^{2\mathrm{i}\varphi(s)}=r(s)\left(1+\mathrm{i}\frac{|t(s)|}{|r(s)|}\right) \label{eq:reful203}
	\end{align}
	with the discretization condition $ \mathrm{e}^{\mathrm{i}k(s)L}=t(s)^*\left(1+\mathrm{i}\frac{|r(s)|}{|t(s)|}\right). $ 
\subsection{Gap equation in the infinite-length limit}
	\indent Now we move to the derivation of the gap equation in the infinite-length limit. Since  $ |u_{\text{PB}}(x,s)|^2+|v_{\text{PB}}(x,s)|^2\xrightarrow{x \to \pm\infty} 2(1+s^{-2})+\text{(oscillating terms)} $, the relation, $ \lim_{L\rightarrow\infty}\frac{1}{L}\int_{-L/2}^{L/2}\left(|u_{\text{PB}}(x,s)|^2+|v_{\text{PB}}(x,s)|^2\right)\mathrm{d}x=2(1+s^{-2}) $, follows. 
	When $ L $ is sufficiently large, the wavenumbers of discretized eigenstates are given by $ k(s)=\frac{2\pi n}{L} $ with an integer $ n $. Therefore the sum is replaced by
	\begin{align}
		\frac{1}{L}\sum_{\substack{\text{s.s.} \\ \epsilon\gtrless 0}}\xrightarrow{L \to \infty}\int_{-\infty}^\infty\frac{\mathrm{d}k}{2\pi}=\pm\int_0^{\pm\infty}\frac{m(1+s^{-2})\mathrm{d}s}{4\pi}
	\end{align}
	Thus, the contribution of positive- (negative-) energy scattering states to the gap equation can be evaluated as
	\begin{align}
		&\frac{1}{L}\sum_{\substack{\text{s.s.} \\ \epsilon\gtrless 0}}\frac{u_{\text{PB}}(x,s)v_{\text{PB}}(x,s)^*}{\frac{1}{L}\int_{-L/2}^{L/2}\left(|u_{\text{PB}}(x,s)|^2+|v_{\text{PB}}(x,s)|^2\right)\mathrm{d}x} \rightarrow \pm\int_0^{\pm\infty}\frac{mu_{\text{PB}}(x,s)v_{\text{PB}}(x,s)^*\mathrm{d}s}{8\pi} \nonumber \\
		&=\pm\int_0^{\pm\infty}\frac{m(u(x,s)v(x,s)^*+r(s)^*u(x,s)^2)\mathrm{d}s}{4\pi} \label{eq:gapssinf01}
	\end{align}
	To obtain the last line, we have used Eq.~(\ref{eq:reful203}) and the relations $\int\mathrm{d}sr(s)v(x,s)^{*2}=\int\mathrm{d}sr(s)^*u(x,s)^2 $ and $ \int\mathrm{d}s\frac{|t(s)|}{|r(s)|}r(s)v(x,s)^{*2}=\int\mathrm{d}s\frac{|t(s)|}{|r(s)|}r(s)^*u(x,s)^2 $, which can be shown using the relations  $ r(s^{-1})=r(s)^* $ and $ s^{-1}v(x,s^{-1})^*=u(x,s) $ valid for real $s$. 
	Using the expression (\ref{eq:gapssinf01}), and following the same discussion in Ref.~\cite{TakahashiNitta}, we obtain the gap equation (\ref{eq:gapeqforLR}). By a similar procedure, we also obtain Eq.~(\ref{eq:gapeqforR}) for a system with only right-movers.
\section{Proof of reflectionless nature}
\subsection{Gel'fand-Levitan-Marchenko equation}
	The Gel'fand-Levitan-Marchenko (GLM) equation, which determines the kernel $ K(x,y) $ of the integral representation for the left Jost solution [Eq.~(S.40)], is given by Eqs.~(S.59)-(S.62). If we rewrite them in the matrix form using the relation (S.43), we obtain
	\begin{gather}
		K(x,y)+F(x,y)+\int_{-\infty}^x\mathrm{d}z K(x,z)F(z,y)=0, \\
		F(x,y):=\frac{m}{4\pi}\int_{-\infty+\mathrm{i}0}^{\infty+\mathrm{i}0}\mathrm{d}s r(s)f_0(x,s)f_0(y,s)^T\sigma_1+\sum_jC_j^2s_jf_0(x,s_j)f_0(y,s_j)^T\sigma_1
	\end{gather}
	with $ f_0(x,s):=\left(\begin{smallmatrix}s^{-1} \\ 1 \ \end{smallmatrix}\right)\mathrm{e}^{-\mathrm{i}k(s)x} $. 
	Following the result of the Riemann-Hilbert approach \cite{AblowitzSegur}, the kernel $ K(x,y) $ can be written as
	\begin{align}
		K(x,y)&=-\frac{m}{4\pi}\int_{-\infty+\mathrm{i}0}^{\infty+\mathrm{i}0}\mathrm{d}sr(s)\frac{f_-(x,s)}{s}f_0(y,s)^T\sigma_1-\sum_j C_j^2s_j\frac{f_-(x,s_j)}{s_j}f_0(y,s_j)^T\sigma_1. \label{eq:RHK01}
	\end{align}
	\indent By substituting Eq.~(\ref{eq:RHK01}) into the integral representation of  $ f_-(x,s) $ [Eq.~(S.40)], and recalling the notations for scattering and bound states introduced in Eq.~(\ref{eq:reful401}), we obtain
	\begin{align}
		\begin{pmatrix} u(x,s) \\ v(x,s) \end{pmatrix}&=\left[\begin{pmatrix}1 \\ s^{-1} \end{pmatrix}+\int_{-\infty-\mathrm{i}0}^{\infty-\mathrm{i}0}\frac{\mathrm{d}\zeta}{2\pi\mathrm{i}} r(\zeta)^*\begin{pmatrix} u(x,\zeta) \\ v(x,\zeta) \end{pmatrix}\frac{\mathrm{e}^{\mathrm{i}k(\zeta)x}}{(1-\zeta s)}\right. \nonumber \\
		&\qquad\qquad\qquad\qquad\qquad\quad \left. +\frac{2\mathrm{i}}{m}\sum_j\begin{pmatrix} f_j(x) \\ s_jf_j(x)^* \end{pmatrix}\frac{e_j(x)}{s_j-s}\right]\mathrm{e}^{\mathrm{i}k(s)x} \label{eq:RHscat001} \\
		\intertext{and}
		\begin{pmatrix} f_j(x) \\ s_jf_j(x)^* \end{pmatrix}&=-\left[ \begin{pmatrix} 1 \\ s_j \end{pmatrix}+\int_{-\infty-\mathrm{i}0}^{\infty-\mathrm{i}0}\frac{\mathrm{d}\zeta}{2\pi\mathrm{i}} r(\zeta)^*\begin{pmatrix} u(x,\zeta) \\ v(x,\zeta) \end{pmatrix}\frac{\mathrm{e}^{\mathrm{i}k(\zeta)x}}{(1-\zeta s_j^{-1})}\right. \nonumber \\
		&\qquad\qquad\qquad\qquad\qquad \left. +\frac{2\mathrm{i}}{m}\sum_l\begin{pmatrix}f_l(x) \\ s_lf_l(x)^*\end{pmatrix}\frac{e_l(x)}{s_l-s_j^{-1}} \right]e_j(x). \label{eq:RHbound001}
	\end{align}
	Here, we have used the relation $ r(\zeta)^*=r(\zeta^{-1}) $ for real $ \zeta $. Note that Eqs.~(\ref{eq:RHscat001}) and (\ref{eq:RHbound001}) provide a closed set of equations for left Jost solutions expressed without using the kernel $ K(x,y) $. 
	Using the relation $ \Delta(x)=m+2\mathrm{i}K_{12}(x,x) $ (Prop. 2 of Supplemental Material of Ref.~\cite{TakahashiNitta}) and Eq.~(\ref{eq:RHK01}) with changing variable $ s=\zeta^{-1} $, we obtain
	\begin{align}
		\Delta(x)=m+m\int_{-\infty-\mathrm{i}0}^{\infty-\mathrm{i}0}\frac{\mathrm{d}\zeta}{2\pi\mathrm{i}} r(\zeta)^*u(x,\zeta)\mathrm{e}^{\mathrm{i}k(\zeta)x}+2\mathrm{i}\sum_js_j^{-1}e_j(x)f_j(x). \label{eq:RHgap01}
	\end{align}
	When $ r(\zeta)\equiv 0 $, Eqs.~(\ref{eq:RHscat001}), (\ref{eq:RHbound001}) and (\ref{eq:RHgap01}) reduce to Eqs. (L.19), (L.13), and (L.14).
\subsection{Proof of the reflectionless nature of self-consistent multi-soliton solutions}
	Using the first components of Eqs.~(\ref{eq:RHscat001}) and (\ref{eq:RHbound001}) and Eq.~(\ref{eq:RHgap01}), after a little long calculation, we obtain
	\begin{align}
		&\frac{m}{2}u(x,s)v(x,s)^*-\frac{\Delta(x)}{2s}=\frac{m}{2}u(x,s)\frac{u(x,s^{-1})}{s}-\frac{\Delta(x)}{2s} \nonumber \\
		&=-2\sum_js_j^{-1}f_j(x)^2\frac{\sin\theta_j}{|s-s_j|^2}+\frac{m}{2}\int_{-\infty-\mathrm{i}0}^{\infty-\mathrm{i}0}\frac{\mathrm{d}\zeta}{2\pi\mathrm{i}}r(\zeta)^*u(x,\zeta)^2\frac{1-\zeta^2}{(s-\zeta)(1-s\zeta)} \label{eq:afterlong01}
	\end{align}
	for real $ s $. Here $ \theta_j $ is defined by $ s_j=\mathrm{e}^{\mathrm{i}\theta_j} $ (See Ref.~\cite{TakahashiNitta} for more detail). 
	Using Eq.~(\ref{eq:afterlong01}) and the formulae
	\begin{align}
		\int\frac{\sin\theta_j\mathrm{d}s}{|s-s_j|^2}=\tan^{-1}\left[ \frac{s-\cos\theta_j}{\sin\theta_j} \right],\quad \int\frac{(1-\zeta^2)\mathrm{d}s}{(s-\zeta)(1-\zeta s)}=\log\frac{s-\zeta}{1-\zeta s},
	\end{align}
	we obtain
	\begin{align}
		&(\text{R.H.S. of Eq.~(\ref{eq:gapeqforLR})})\nonumber \\
		&=\sum_js_j^{-1}f_j(x)^2\left[ \nu_j-\frac{2\theta_j-\pi}{\pi} \right]+\frac{m}{2}\int_{-\infty}^\infty\frac{\mathrm{d}s}{2\pi}r(s)^*u(x,s)^2\left[ \frac{\log s^2}{\mathrm{i}\pi}-\operatorname{sgn}s \right].
	\end{align}
	Because of linear independence of  $ f_j(x) $'s and $ u(x,s) $'s, we finally obtain
	\begin{align}
		\nu_j=\frac{2\theta_j-\pi}{\pi},\quad r(s)= 0,
	\end{align}
	for $ j=1,\dots, n $ and any real $ s $, and we thus conclude the reflectionless property $ r(s)=0 $. On the other hand, for the system with only right-movers, the gap equation [Eq.~(\ref{eq:gapeqforR})] becomes
	\begin{align}
		&(\text{R.H.S. of Eq.~(\ref{eq:gapeqforR})})\nonumber \\
		&=\sum_js_j^{-1}f_j(x)^2\left[ \nu_j-\frac{\theta_j}{\pi} \right]+\frac{m}{2}\int_{-\infty}^\infty\frac{\mathrm{d}s}{2\pi}r(s)^*u(x,s)^2\left[ \frac{\log s^2}{2\mathrm{i}\pi}+H(-s) \right],
	\end{align}
	where $ H(s) $ is the Heaviside step function. Thus the self-consistent condition is given by
	\begin{align}
		\nu_j=\frac{\theta_j}{\pi},\quad r(s)=0,
	\end{align}
	for $ j=1,\dots, n $ and any real $ s $. The result $ \nu_j=\theta_j/\pi $ is consistent with the recent work\cite{DunneThies}.
\section{Summary}
	In this paper, we have proved that the self-consistent soliton solutions must have reflectionless potentials. Combining this paper and our previous work\cite{TakahashiNitta}, we can provide the fully self-contained derivation of static self-consistent solutions under the uniform boundary condition. One of future problems is to generalize the solutions with modulated background, using the solution recently obtained in Ref.~\cite{TakahashiarXiv}.

\begin{acknowledgements}
The work of M.~N. is supported in part by KAKENHI (No. 25400268 and 25103720).
\end{acknowledgements}


\end{document}